# A Flexible Design for Optimization of Hardware Architecture in Distributed Arithmetic based FIR Filters

Fazel Sharifi, Saba Amanollahi, Mohammad Amin Taherkhani and Omid Hashemipour
*Faculty of Electrical and Computer Engineering, Shahid Bebehshti University,G.C., Tehran, Iran*
{f_sharifi, s_amanollahi, m_taherkhani, hashemipour}@sbu.ac.ir

**Abstract**

*FIR filters are used in many performance/power critical applications such as mobile communication devices, analogue to digital converters and digital signal processing applications. Design of appropriate FIR filters usually causes the order of filter to be increased. Synthesis and tape-out of high-order FIR filters with reasonable delay, area and power has become an important challenge for hardware designers. In many cases the complexity of high-order filters causes the constraints of the total design could not be satisfied. In this paper, efficient hardware architecture is proposed for distributed arithmetic (DA) based FIR filters. The architecture is based on optimized combination of Look-up Tables (LUTs) and compressors. The optimized system level solution is obtained from a set of dynamic-programming optimization algorithms. The experiments show the proposed design reduced the delay cost between 16%-62.5% in comparison of previous optimized structures for DA-based architectures.*

## 1. Introduction

Nowadays, FIR filters, regarding to their superior properties such as stability and high reliability in digital signal processing, have had many important and widespread applications. This kind of digital filters are applied to an extensive form in many areas such as image processing, radio communication and high technology devices. One of the important applications of FIR filters is in analog to digital converters [1][2]. Also FIR filters are applied in read channel of disc drives known as PRML [3] in addition to wideband receivers in wireless communication devices [4].

Generally in an *N*-order FIR filter with $coef[i] \;\forall i \in [0, N-1]$, the output *y[t]* is calculated according to the current input *x[t]* and previous inputs $x[t-i] \;\forall i \in [1, N-1]$ by equation (1):

$$y[t] = \sum_{i=0}^{N-1} coef[i]x[t-i]$$
$$= coef[0]x[t] + coef[1]x[t-1] + \ldots + coef[N-1]x[t-N+1] \quad (1)$$

As it is shown in equation (1), in each step for calculation of *y[t]* as the output of *N*-order filter, *N* additions and *N* multiplications are required.

Although the FIR filters are more stable compared to IIR filters, their circuits are very complicated due to necessity of designing high order FIR filters in real application. By increasing in order of filter, the circuit is become more complicated and this is one of the most important challenges in front of designing these types of filters. This complexity sometimes causes desired design does not meet area, timing and power constraints.

In this paper a compound architecture for FIR filters is proposed. Considered modules in the proposed architecture are optimized by efficient algorithms and final architecture will be extracted.

The rest of this paper includes following sections. In section 2, architectures of FIR filter with related works are reviewed. In section 3 the proposed architecture is introduced. Optimization algorithms for construction of efficient hardware architecture are introduced in section 4. Experimental results are shown in 5 and finally section 6 concludes the paper and future works are considered.

## 2. A review of related architectures

For implementation of FIR filter structures, hardware architectures based on multiply and add (MAC) and distributed arithmetic (DA) are known as the main classes of FIR filter architectures.

In MAC-based architectures, computation of the desired output is done directly and it is based on multiplication and addition. To improve the performance of MAC-based architectures, some researches focused on design of filters based on Residue Number system (RNS) [5]-[7]. In these designs, processing overhead and hardware cost of binary to RNS conversion should be considered.

Complexity of classic multiplication and addition causes a lot of researches tend to change in filter architecture and be done based on DA, so that without using multipliers, the idea of pre-computing and storing the required values has been exploited.

The classic method of distributed computing works based on changing the form of required computations in equation (1) and rewriting it in new computation forms. For this purpose it assumes, the states $x[k] \;\forall k \in [t-N+1, t]$ in binary standard format (2's complement) and has been

scaled ($|x[k]| < 1$) and it can be shown by the following equation:

$$x[k] = -x_{B-1}[k] + \sum_{j=0}^{B-2} x_j[k] 2^{-j} \quad (2)$$

Substituting equation (2) in equation (1) results to:

$$y[t] = \sum_{i=0}^{N-1} coef[i](-x_{B-1}[t-i]) + \sum_{j=0}^{B-1} 2^j \times (\sum_{i=0}^{N-1} coef[i].x_j[t-i]) \quad (3)$$

A look up table could be used with *N*-bit input address to preserve the value of $\sum_{i=0}^{N-1} coef[i].x_j[t-i]$. In This condition, size of LUT size (without applying optimization) would be $2^N \times (C + \log_2^N)$, where *N* is the order of filter *C* is the length of coefficients in binary form. As the equation 3 shows, a combination of sum of coefficients can be calculated in advance and be stored in look up tables.

The main problem of look-up tables is the complexity of table size which grows exponentially by increasing of filter order. A method have been proposed in [9] to reduce (and ultimately eliminate) the size of required look-up table. But in the proposed LUT-less architecture, the delay of adders and multiplexers (MUX) is not considered and therefore the solution is not efficient for performance-critical applications. In [10] another DA-based FIR filter has been presented that is suitable for FPGA platforms with 4-input look-up tables.

## 3. Proposed architecture

As it has been shown in previous section, distributed arithmetic based architectures face off with exponential complexity problem for the size of look up tables. In this section an efficient architecture is presented in order to increase performance of computation part of digital filters. As it is illustrated in Figure 1, main components of proposed architecture are formed based on a compound model with two main layers including look up tables and compressors. As it can be seen in Figure 1, *N* bits are used in first layer of shift register and $k = \sum_{i=1}^{m} k_i$ bits out of these *N* bits are assigned to *M* look up tables, with $k_1 \ldots k_m$ inputs. In the next section, an optimization algorithm for finding efficient structure for look up tables set with *k* bits input address is introduced.

Remaining bits (*N-k*) are used as selectors in 2-1 multiplexers (for every bit) and totally $C \times (N - k)$ multiplexers are required. If the selector of $i^{th}$ multiplexer becomes 1 the related coefficient will be added to compressor part.

*M* outputs of look up tables with *N-k* outputs of multiplexers are used as inputs of *N-k+m:2* compressor. The functionality of the compressor is shown in Figure 2. Extraction of an efficient *N-k+m:2* compressor is described in the next section. After compression, a CLA adder is used for final summation.

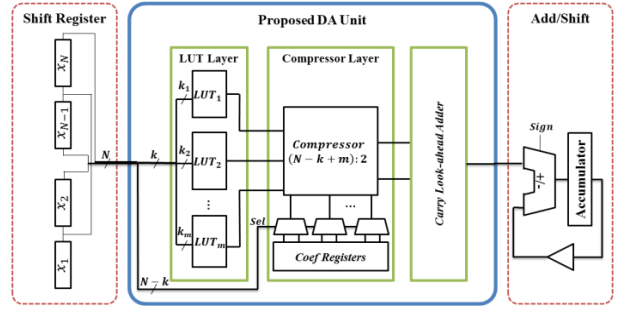

**Figure 1. Proposed Architecture for Distributed Arithmetic Unit**

It should be considered that look up table's layers and compressors can be used alone without each other. In the other hand optimized architecture can work without compressor (Partitioned-LUT) or without look up tables (LUT-less). In the next section, three algorithms have been proposed to identify the optimum architecture.

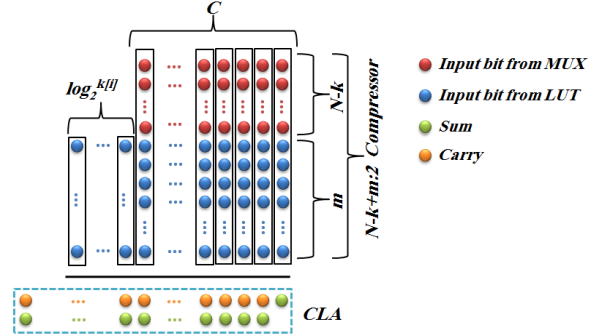

**Figure 2. Functionality of the designed Compressor Layer**

## 4. Architecture extraction algorithms

As was noted in the previous section, the proposed approach provides possibility of choosing suitable hardware architecture based on compound structure sets in a flexible way. For finding efficient architecture the following parameters should be determined precisely:

- Optimized structure of LUT set based on input parameter (*k*): In this section $k_i$ and *m* values are chosen somehow LUT size with *k* bit input address is optimized. The proposed algorithm is presented in section 4.1.

- Optimized compressor structure based on input bits set (*h*): In this section optimized compressor structure is extracted based on small-optimized compressors. The proposed algorithm is presented in section 4.2.

- Final optimized architecture: In this section, according to parameter values for LUTs and compressors, number of input addresses for the LUT layer and number of selector bits for compressor layer are chosen in a way which the final architecture is optimized. Final optimization algorithm is proposed in section 4.3.

Optimization could be done based on following parameters:

- Gate latency: in all optimizations process delay of XOR gate is considered as the delay unit. This parameter is shown by (2ΔG).
- Power consumption: optimization for power consumption is based on minimum number of resources and could be determined by power consumption unit of XOR gate.
- Power-delay product (PDP): the proposed algorithms could be extended to optimize PDP parameter based on previous parameters.

### 4.1. Optimization of LUT Layer

In this section, technique has been proposed for partitioning of LUTs. In this work the LUT layer with $k$ bits for input addresses and $m$ outputs is partitioned into $m$ basic LUTs based on computation of delay/power and power delay product (PDP) parameters in gate level. LUT models based on Decoder-Memory are exploited for description of architecture in the LUT layer [14].

The structure of a basic LUT with $k_i$ input address (for preserving the summation of $k_i$ coefficients) contains a $k_i$ to $2^{k_i}$ decoder and a $2^{k_i}$ words memory with $C + [log_2^{k_i}]$ length. Suppose, delay of this LUT is notated as $d_{lut\ k[i]}$, and its power consumption and power delay product (PDP) are respectively notated as $p_{lut\ k[i]}$ and $pdp_{lut\ k[i]}$. Therefore for a LUT layer with the $k$ inputs and $m$ outputs, the structure with optimized delay ($D(LUT_{k,m})$) or optimized power ($P(LUT_{k,m})$) or optimized PDP ($PD(LUT_{k,m})$) could be obtained from the dynamic programming Algorithm 1 with $O(n^2)$ complexity.

### 4.2. Optimization of Compressor Layer

In this section a method is proposed for auto construction of the $h:2$ optimized compressor based on delay, power and PDP parameters. Creating large input compressors are carried out by using of optimized conventional and unconventional compressors [11]-[13].

Therefore, for creating an optimized large input compressor ($h:2$), set of $F$ basic compressors $Comp_i...Comp_F$ with compression levels $(I_{Comp[k]} : O_{Comp[k]}) \forall k \in [1..F]$ are used. Unconventional compressors have some carry in and carry out bits. However, these carry bits are created in such a way in compressor which there are not carry propagation. Suppose delay, power and PDP of basic compressor $k$ are $d_{Comp[k]}, p_{Comp[k]}, pdp_{Comp[k]}$ respectively.

---

*OptimizeLUT (Address Bits: k, Number of LUTs :m):*
1. $D(LUT_{i,1}) \leftarrow d_{lut[i]};$      $\forall i \in [1..k]$
2. $P(LUT_{i,1}) \leftarrow p_{lut[i]};$      $\forall i \in [1..k]$
3. $PD(LUT_{i,1}) \leftarrow pd_{lut[i]};$      $\forall i \in [1..k]$
4. **for** (i=2; i <= k; i++)
5.    **for** (j=2; j <= m; j++)
6.      $D(LUT_{i,j}) \leftarrow min_u\{max\{d_{lut[u]}, D(LUT_{i-u,j-1})\}\};$
7.      $P(LUT_{i,j}) \leftarrow min_w\{p_{lut[w]} + P(LUT_{i-w,j-1})\};$
8.      $PD(LUT_{i,j}) \leftarrow min\{D(LUT_{i,j}).P_u(LUT_{i,j}), D_w(LUT_{i,j}).P(LUT_{i,j})\};$
9.    **end for**
10. **end for**
11. **return** $D(LUT_{k,m}), P(LUT_{k,m}), PD(LUT_{k,m});$

**Algorithm 1. Proposed Algorithm for Optimization of LUT layer**

The problem of finding minimum Delay ($D_{h,2}(k)$-which describes minimum delay of h:2 compressor with basic compressors $Comp_1...Comp_k$) or minimum Power ($P_{h,2}(k)$) or minimum PDP ($PDP_{h,2}(k)$) could be followed by two different configurations. In one configuration, the $Comp_k$ is not used and therefore the best solution may be gathered from previous calculations $D_{h,2}(k-1), P_{h,2}(k-1), PDP_{h,2}(k-1)$. But the other way is usage of $Comp_i$. In this condition, the compressor is divided into 3 sub-modules as shown in Figure 3. These modules are the basic compressor $Comp_k$ and two compound compressors $h-I_{Comp[k]}:g$ and $O_{Comp[k]}+g:2$. The optimum solution is obtained from minimum of these two configurations. The proposed dynamic programming approach with polynomial complexity is shown in Algorithm 2.

### 4.3. Extraction of final solution

In this step, based on optimization results of the LUT layer and the compressor layer, the final architecture of the proposed DA unit is extracted. In other word the value for $m$ and $k$ is determined by using the Algorithm 3.

Based on the main criteria for the designer, the algorithm could present separately the optimized solution for delay, power or PDP parameters. As shown in Algorithm 3, the cost function could be any arbitrary parameters Delay, Power or PDP returned from *OptimizedLUT* and *OptimizedComp* Algorithms.

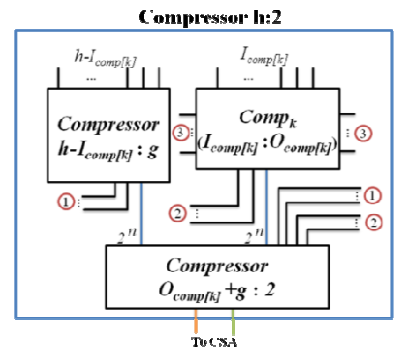

**Figure 3. A configuration of h:2 compressor exploiting the basic compressor $Comp_k$**

```
OptimizeComp (CompLevel: h):
1.  F← Number of basic Compressors
2.  D_{i,j}(0)←∞; P_{i,j}(0)←∞; PDP_{i,j}(0)←∞; ∀i ∈ [1..h], j < i
3.  D_{i,j}(k)←0; P_{i,j}(k)←0; PDP_{i,j}(k)←0; ∀k ∈ [1..f], i ∈ {1,2}, j < i
4.  while (k < F)
5.    for (i=3; i <= h; i++)
6.      for (j=1; j < i; j++)
7.        if((i:j) = (I_{Comp[k]}: O_{Comp[k]}))
8.          D_{i,j}(k) ← d_{Comp[k]};
9.          P_{i,j}(k) ← p_{Comp[k]};
10.         PDP_{i,j}(k)← pdp_{Comp[k]}
11.       else
12.         D_{min}←min_u{max{D_{i-I Comp[k],u}(k), d_{Comp[k]}}+D_{O Comp[k]+u,2}(k)};
13.         P_{min}←min_w{P_{i-I Comp[k],u}(k) + P_{O Comp[k]+u,2}(k)+ p_{Comp[k]} };
14.         TP←ComputePower(u); TD←ComputeDelay(w);
15.         D_{i,j}(k) ← min{D_{i,j}(k-1), D_{min}};
16.         P_{i,j}(k) ← min{P_{i,j}(k-1), P_{min}};
17.         PDP_{i,j}(k) ← min{D_{i,j}(k-1). P_{i,j}(k-1), D_{min}.TP, P_{min}.TD};
18.     end for
19.   end for
20.   k++;
21. end while
22. return  D_{h,2}(F), P_{h,2}(F), PDP_{h,2}(F);
```

**Algorithm 2. Optimization of Compressor Layer**

```
OptimizeArch (Filter Order: N):
1.  OptSolution←∞;
2.  Select cost from {Delay | Power | PDP}
3.  for (i=1; i <= N; i++)
4.    for (j=1; j <=i; j++)
5.      ArchCost = cost(OptimizeLUT(i,j), OptimizeComp(N-i+j));
6.      if (ArchCost<OptSolution)
7.        OptSolution←ArchCost;
8.    end for
9.  end for
10. return OptSolution
```

**Algorithm 3. Extraction of optimized architecture**

## 5. Implementation and Experiments

Implementation has performed in two sections. In first section, hardware description of all basic components has been implemented. Verilog implementation of basic optimized compressors includes 3:2, 4:2, 5:2, 6:2, 7:2 and 9:2 compressors [11]-[13]. The other regular compressors such as 7:3 or 15:4 could be constructed from these basic components. The implementation of basic LUTs was based on memory model in CACTI 5.1[14]. In the second section of this phase, the optimized architecture algorithms have been implemented in 9 source and header files to present the optimized structure of DA unit.

The real coefficients have been extracted from Filter Design and Analysis Tool (FDATool) [15]. The coefficients are produced for implementation of the sample filters listed in Table 1. As shown in the table based on the supplied criteria, the order of filter is determined from 8 to 143. According to our requirements in application of designing digital part of ADCs, the frequency of sampling ($F_s$) and length of inputs ($B$) have been set to a 40MHz and 3 bits respectively. In addition, in all design $C$ (length of coefficients) is set to 16 bits.

**Table 1 Specification of analyzed FIR filters**

| Filter Order | $F_S$ (MHz) | $F_{Pass}/F_{Stop}$ (MHz) | $A_{Pass}/A_{Stop}$ (dB) |
|---|---|---|---|
| 8 | 40 | 1.2/3.8 | 3/20 |
| 18 | 40 | 1.2/3.8 | 3/40 |
| 31 | 40 | 1.6/3.0 | 3/40 |
| 72 | 40 | 2.2/2.8 | 3/40 |
| 108 | 40 | 2.2/2.8 | 3/60 |
| 143 | 40 | 2.2/2.8 | 3/80 |

The estimated delay based on (Gate delay-ΔG) for Distributed Arithmetic unit of LUT, LUT-less [9] and proposed architectures is shown in Figure 4. LUT-less architecture in [9] is implemented by full compressors instead of regular adders. As shown in the figure delay of the proposed architecture is 16% (for 8-order filter) to 62.5% (for 143-order filter) less delay in comparison of LUT-less architecture.

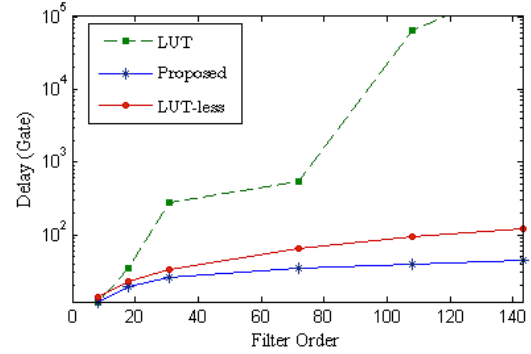

**Figure 4 Estimated Delay of FIR filters for DA-based architectures**

## 6. Conclusion and Future Works

According to design constraints in high-order FIR filters, in this work the following contributions were presented:

- Compound hardware architecture was proposed for distributed arithmetic based FIR filters. The architecture exploits the benefits of pre-served summation in optimized LUTs and improves the speed of addition by using high efficient compressors.
- A dynamic programming algorithm was proposed with polynomial complexity to find the optimized structure of compressors.
- A dynamic programming algorithm was proposed to find the best solution for LUT partitioning.
- The final optimized architecture could be extracted from third-proposed algorithm.

For the future works, we will attempt to extend the tool which is capable for automatic generation of HDL code for the optimized extracted architecture.

**Acknowledgment**

The authors would like to acknowledge members of Micro Electronic Lab in Faculty of Electrical and Computer Engineering in Shahid Beheshti University (SBU) specially, Professor K. Navi, S. Z. Reyhani and Adel Hosseiny.